\documentclass[prl,twocolumn,superscriptaddress]{revtex4-1} 
\usepackage{times}
\usepackage{amsmath,amssymb,mathrsfs}
\usepackage{graphicx}
\usepackage{hyperref}
\usepackage{paralist}
\usepackage{ifthen}
\usepackage{dsfont}
\begin{document}

\newcommand{\FourHe}{\textsuperscript{4}He}
\newcommand{\calV}{{\cal V}}

\title{Quantum Critical Phenomena of \textsuperscript{4}He in Nanoporous Media}

\author{Thomas Eggel}
\author{Masaki Oshikawa}
\affiliation{Institute for Solid State Physics, University of Tokyo,
Kashiwa 277-8581 Japan}

\author{Keiya Shirahama}
\affiliation{Department of Physics, Keio University, Yokohama 223-8522
Japan}

\date{\today}

\begin{abstract}
The superfluid transition in liquid \FourHe\ filled in
Gelsil glass observed in recent experiments is discussed
in the framework of quantum critical phenomena.
We show that quantum fluctuations of phase are indeed important
at the experimentally studied temperature range
owing to the small pore size of Gelsil,
in contrast to \FourHe\ filled in previously studied porous media
such as Vycor glass.
As a consequence of an effective particle-hole symmetry,
the quantum critical phenomena of the system
are described by the 4D XY universality class,
except at very low temperatures.
The simple scaling agrees with
the experimental data remarkably well.
\end{abstract}

\maketitle

{\em Introduction--}
Superfluidity is one of the most impressive macroscopic quantum
phenomena that can be experimentally observed.
Since liquid \textsuperscript{4}He is a very clean system and the U(1) symmetry
of the quantum phase is exact, its superfluid transition
is also an ideal case to study phase transitions.
In fact, much of the most precise experimental estimates
of critical exponents is obtained from the superfluid
transition of \textsuperscript{4}He \cite{space, kleinert}.

In bulk liquid \textsuperscript{4}He, the second-order superfluid
phase transition occurs at finite temperature only.
On the other hand, the properties of liquid \textsuperscript{4}He can be 
changed by confining \textsuperscript{4}He in porous media.
For example, \FourHe\ in Vycor glass exhibits a quantum phase
transition as a function of film thickness, when \FourHe\
forms a two-dimensional film on the interior
surface of the pores~\cite{crowell}. However, the physics might be complicated due to the crossover
from two-dimensional to three-dimensional behaviors.
When \FourHe\ is filled in Vycor pores, the superfluid
transition remains at finite temperature
(see \cite{shira_jpsj} and references therein).
Recently, however, by using Gelsil glass which has nanopores of
2.5 nm mean nominal diameter, considerably smaller than
comparable substrates used in the past, 
the superfluid transition temperature
is suppressed to zero under an applied pressure~\cite{YSS}.
This implies the existence of a quantum critical point
at zero temperature.

The non-superfluid phase close to the quantum critical
point exhibits rather abnormal properties.
This also seems to be closely related to the pseudogap regime
in high-$T_c$ superconductivity~\cite{LNW}. 
The \FourHe\ in nanoporous media is in several ways
(including the absence of fermionic degrees of freedom)
simpler, and might provide a useful insight into
the physics of more complex systems such
as the high-$T_c$ superconductors.

In this Letter, 
we discuss the quantum critical phenomena
of the superfluid transition in
liquid \FourHe\ in Gelsil glass.
We determine the value of the interaction strength $V$ in the
Bose-Hubbard model for liquid \FourHe\ in porous media.
Although the strongly interacting and dense quantum liquid nature
of \FourHe\ makes a microscopic theoretical
calculation of the effective parameter $V$ difficult,
we show that it can be derived from macroscopic
properties of liquid \FourHe.
We find that, although the interaction parameter is about 20 mK for
the Vycor system, it is as large as about 1 K for
the Gelsil system.
This explains the fact that the quantum critical
phenomena are visible for the experimentally studied temperature range
only in the Gelsil system.

The Bose-Hubbard model with disorder has been a subject
of extensive theoretical study. Its quantum critical phenomena
are highly nontrivial, and much of the problem still remains open.
Nevertheless, we argue that 
the particle-hole symmetry breaking due to chemical potential,
which drives the system away from the 4D XY universality class,
is practically negligible in the present system. 
As a consequence, for a large part of the phase diagram,
the quantum critical phenomena observed experimentally
in \FourHe\ in Gelsil glass
can be understood with the simple non-random 4 dimensional XY (4D XY)
universality class~\cite{*[{For related discussions of the superfluid density in quantum critical phenomena in high $T_c$ superconductivity see }][{}] franz, *lemberger}.

We demonstrate that the simple scaling based on the
non-random 4D XY universality class indeed agrees
quite well with the experimental data on
liquid \FourHe\ in Gelsil glass, except in the small
region close to the quantum critical point, where a
crossover to a different universality class is observed.

{\em Setup--}
The Gelsil glass may be modeled by many highly interconnected pores,
each of which
can contain a number of \textsuperscript{4}He atoms.
While the size of the pores is randomly distributed, there
is a typical pore diameter for a given sample.

One of the most outstanding experimental findings on the
system is a rounded peak in the specific heat at temperature $T_B$
somewhat lower than the superfluid transition temperature
$T_\lambda$ in the bulk.
In Ref.~\cite{shira_jpsj} it was proposed
that the specific heat peak should be interpreted as
the formation of localized BEC (LBEC) within each pore. 

In fact, here we demonstrate 
that the peak temperature $T_B$ can be indeed understood as
a rounded lambda transition temperature in a finite-size system.
Standard scaling theory gives
\begin{equation}
 \Delta T \equiv T_\lambda - T_B(l) \propto C l^{-1/\nu_{3D}}, 
\label{eq:Tm}
\end{equation}
where $l$ is the linear pore size, and $\nu_{3D} \sim 0.67$
is the correlation length exponent for the $\lambda$ transition. 

\begin{figure}[h]
   \centering
\includegraphics[width=0.35 \textwidth]{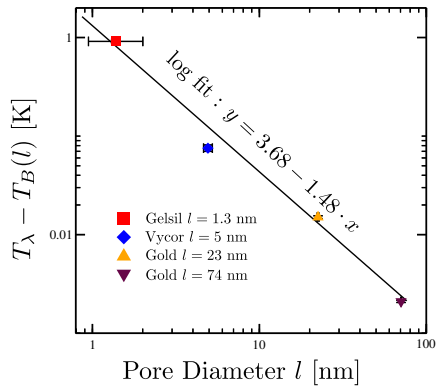}
\caption{
(color online)
The condensation temperature shift $T_{\lambda} - T_B$
as a function of the effective pore size,
which supports the scaling~\protect\eqref{eq:Tm}.
}
\label{pic:shift}
\end{figure}

In two samples of nanoporous glasses with different pore sizes, Gelsil and Vycor,
different values of $\Delta T$ were observed~\cite{YSS,joseph}.
Moreover, a similar shift was also observed
in porous Gold samples with two different pore sizes~\cite{yoon}.
The four values of $\Delta T$ are plotted as a function
of the pore size $l$ in Fig.~\ref{pic:shift}.
Here we use the effective pore size, obtained by
subtracting the inert layer thickness from the
nominal pore diameter.
Although the data were taken for different porous materials
at different pressures, 
the agreement with the scaling~\eqref{eq:Tm} is remarkable.
$\nu_{3D}=1/1.48 = 0.676$ estimated from the fit
is consistent with the known value $\nu_{3D} \sim 0.67$.
In addition to establishing the localized BEC picture,
this result implies that liquid \FourHe\ in the nanopores
does inherit bulk properties.

Since the number of atoms is finite, \textsuperscript{4}He
in a single pore cannot be a superfluid in the true thermodynamic sense.
In fact, the superfluid transition temperature is lower than
 $T_B$. However, below $T_B$, liquid \FourHe\ in each pore
can have an approximately defined quantum phase which
is a characteristic of a condensate. 
On the other hand, superfluidity as a macroscopic phenomenon
only occurs when, by the hopping of atoms between pores,
global phase coherence emerges over the entire system.
This occurs at a temperature $T_c$ which is lower than $T_B$.

The system may be represented by
the (disordered) Bose-Hubbard model~\cite{FWGF}
\begin{equation}
 \mathcal{H}_{\mathrm{BH}} =
\sum_i \left( \frac{V_i}{2} {n_i}^2 - \mu_i n_i \right)
- \sum_{\langle i,j \rangle} \left(
t_{ij} a^{\dagger}_i a_j + \mbox{h.c.}
\right),
\label{eq:BH}
\end{equation}
where each site corresponds to a pore containing a local condensate,
connected to neighboring sites by random hopping
$t_{ij}$.
Every pore has a random \emph{chemical potential} $\mu_i$ and
a random \emph{charging energy} $V_i$.
The indices $i,j$ label the pores and $n_i$
represents the number of \textsuperscript{4}He atoms in
the $i$-th pore.
The charging energy and the chemical potential can be combined as
\begin{equation}
\dfrac{V_i}{2} {n_i}^2- \mu_i n_i  =
\dfrac{V_i}{2} ( n_i - \bar{n}_i)^2 + \mbox{const.},
\end{equation}
where $ \bar{n}_i = \mu_i/V_i$.
The quantum phase $\theta_i$ at site $i$ can be
regarded as canonically conjugate
to the number operator $n_i$,
namely they satisfy the canonical commutation relations
$ [ \theta_j, n_k ] = i \delta_{jk}$,
implying that the phase and atom number in each pore
obey the uncertainty relation.
The ``charging energy'' $V_i$ tends to fix $n_i$ to $\bar{n}_i$,
thus introducing quantum fluctuations of the phase $\theta_i$
due to the uncertainty relations.

\emph{Quantum Fluctuations--}
It is important to estimate
the typical value of the ``charging energy'' $V_i$.
In the application of the Bose-Hubbard model to superconductors,
$V_i$ represents the charging energy due to Coulomb
repulsion. In contrast, in the present case of
neutral liquid \FourHe, there is no Coulomb repulsion.
Nevertheless, putting an extra atom to,
or removing an atom from, a pore in the groundstate
should lead to an increase in energy.
This can be related to the finite compressibility of the liquid in the pore.
In fact, we can estimate $V_i$ as follows:
\begin{equation}
 \frac{1}{V_i} = \frac{\partial \bar{n}_i}{\partial \mu_i}
= \calV_i \nu^2 \kappa,
\label{eq:interaction}
\end{equation}
where $\calV_i$ is the (effective) volume of the pore,
and $\nu$ and $\kappa$ are the number density
and the compressibility of liquid \FourHe, respectively.
We approximate $\nu$ and $\kappa$ of \FourHe\ inside the pores
by their values in the bulk:
$\nu \sim 2.1\times10^{28} \; m^{-3},
\kappa \sim 10^{-7} \; \mbox{Pa}^{-1}$.

Then, assuming that a typical pore is a sphere with
effective diameter $2 R_{\mbox{\scriptsize eff}} =1.3$ nm
(subtracting the thickness 0.6 nm of the inert 
layer), we obtain an estimate
for the typical value of $V_i \sim  1.4 \mbox{K} $ for \FourHe\ in Gelsil Glass.
In contrast, for Vycor Glass with effective diameter 5 nm,
the same argument leads to $V_i \sim 0.02 \mbox{K}$.

If $T \gg V_i$,
the quantum fluctuations of the phase may be neglected.
In such a limit, although the phase $\theta_i$ is a quantum-mechanical
degree of freedom, it can be regarded as a classical variable.
On the other hand, when $T \lesssim V_i$, quantum fluctuations
of the phase become important.
Thus, in the temperature range probed in the experiments,
quantum fluctuations would be important for \FourHe\ filled in
Gelsil glass but not in Vycor glass.
This is consistent with the experimental results that
quantum critical phenomena are observed only in Gelsil glass,
when pores are filled by \FourHe .

In the continuum limit,
the (disordered) Bose-Hubbard model~\eqref{eq:BH}
is described by the Lagrangian density in $3+1$ dimensions:
\begin{align}
{\cal L} = & \frac{1}{2} | \nabla \psi |^2
- \frac{1}{2} \psi^* [\partial_\tau - g_0 - \delta g(x)]^2 \psi
+ \frac{1}{2} (r_0 + \delta r(x)) |\psi|^2 
\notag \\
& + \frac{1}{4} u_0 | \psi|^4,
\label{eq:Lag}
\end{align}
where $\psi$ is a complex scalar field, $g_0, r_0$, and $u_0$ are
constants, and $\delta g(x)$ and $\delta r(x)$ are
position-dependent random
variables distributed around zero~\cite{FWGF,weichman_prl,*weichman}.

When $g_0 = \delta g = \delta r = 0 $, this theory reduces to
the standard $\psi^4$ theory with the dynamical critical
exponent $z=1$, and the quantum critical point
is described by the 4D XY universality class.
Since 4 is the upper critical dimension for the XY model, the critical
exponents are given by the mean-field theory.
For example, the correlation length exponent is given by 
$\nu_{4D}=1/2$ ~\footnote{
In this Letter, we shall ignore the expected logarithmic corrections,
which are not visible in the data so far.}.
This limit corresponds to, 
in terms of the original model~\eqref{eq:BH},
the special case where there is no disorder and 
$\bar{n}_i = \mu_i / V_i$'s are exactly an integer.
In particular, $g_0 = \delta g =0$ implies an exact
particle-hole symmetry.

In reality, $\bar{n}_i$ is randomly distributed, breaking
the particle-hole symmetry.
Its effect can be classified into the overall
symmetry breaking $g_0$ and
the local symmetry breaking $\delta g$,
in the continuum theory~\eqref{eq:Lag}.
$g_0$ is generically non-vanishing and is a relevant
perturbation to the 4D XY fixed point.
In fact, even in a system without randomness, $g_0$ drives
the system to a different critical behavior except at
special multicritical points (tips of the Mott lobes).
Nevertheless, here we argue that the effect of $g_0$
is practically negligible for the present system,
partly owing to the randomness.

Let us assume that
the distribution of the pore radius has standard
deviation of, say, $\Delta R = 0.02 \mbox{nm}$,
which would be rather an underestimation.
This translates to the width
\begin{equation}
 \Delta \bar{n} \sim 4 \pi \nu {R_{\mbox{\scriptsize eff}}}^2 \Delta R
\sim 2.2
\label{eq:barnwidth}
\end{equation}
for the distribution of the $\bar{n}_i$.
We assume that the $\bar{n}_i$ follow a Gaussian distribution with average
$\bar{n}_{\mbox{\scriptsize av}}$
and standard deviation $\Delta \bar{n}$.
Since the effect of the particle-hole symmetry breaking is a
periodic function of $\bar{n}_i$, it may be estimated by
$\sin{2\pi \bar{n}_i}$.
The effective overall particle-hole symmetry breaking $g_0$
then reads
\begin{equation}
 \int \sin{(2\pi x)}
 \frac{e^{-\frac{(x-\bar{n}_{\mbox{\scriptsize av}})^2}{2 (\Delta
 \bar{n})^2}}}{\sqrt{2 \pi} \Delta \bar{n}} 
  \; dx = \sin{(2 \pi \bar{n}_{\mbox{\scriptsize av}})}
e^{-2 \pi^2 (\Delta \bar{n})^2} .
\end{equation}
The first factor $ \sin{(2 \pi \bar{n}_{\mbox{\scriptsize av}})}$ just
represents the particle-hole symmetry breaking for the
average value $\bar{n}_{\mbox{ \scriptsize av}}$, which is
generically non-vanishing.
The second factor $e^{-2 \pi^2 (\Delta \bar{n})^2}$ shows the
suppression of the symmetry breaking by the random distribution.

For the width given in eq.~\eqref{eq:barnwidth},
the suppression is in fact about $10^{-43}$.
Thus, in the realistic temperature range,
the overall particle-hole symmetry breaking $g_0$
is negligible.
That is, in the present system, each pore contains enough 
particles so that the distinction between the particle
and hole becomes unimportant.
A similar discussion was given for the asymptotic
low-energy behavior of disordered bosons ~\cite{weichman_prl,*weichman}.
Here the situation is somewhat different in that
the smallness of $g_0$ is
not a consequence of the RG transformation, but
is rather of microscopic origin.

The random particle-hole symmetry breaking $\delta g$
is an irrelevant perturbation to
the 4D XY (Gaussian) fixed point ~\cite{weichman_prl,*weichman}.
Although the presence of $\delta g$ is believed to be
important to determine the eventual fate of the RG flow,
it can be ignored in the neighborhood of the 4D XY fixed point.
Thus $r_0$ and $\delta r$ are the only remaining relevant
parameters around the 4D XY fixed point.
This theory represents the critical behavior of
a classical XY model in 4 dimensions.
In the absence of the random $\delta r$,
the quantum phase transition belongs to the 4D XY
universality class.
$r_0$ corresponds to the temperature of
the classical XY model, and is the control parameter 
for the quantum phase transition.
It is thus identified with the pressure $p$ of
\FourHe\ in the system.

The random term
$\delta r$ corresponds to disorder in the classical XY model.
We assume that the disorder is spatially uncorrelated.
However, in the mapping to the classical XY model,
the disorder is completely correlated in the imaginary
time direction.
This class of disorder turns out to be a relevant perturbation
to the 4D XY fixed point, driving the system to a different
fixed point ~\cite{cardy}.
This is in contrast to the effect of disorder on
the finite temperature superfluid transition,
where the uncorrelated disorder is an irrelevant perturbation
according to the Harris' criterion ~\cite{harris,Kiewiet-He-Vycor-PRL1975}.

The new fixed point induced by $\delta r$,
corresponding to the four-dimensional XY model with
disorder correlated in the imaginary time direction,
is sometimes called the random rod fixed point.
At this random rod fixed point, the random part of
the chemical potential $\delta g$ is believed
to be relevant, eventually driving the system
to yet another (disordered boson) fixed point ~\cite{weichman_prl,*weichman}.
However, if $\delta r$ is small, the system
may be described by the 4D XY fixed point
and the crossover away from it due to $\delta r$,
down to a certain temperature.

We argue that this is indeed the case
in the present system of \FourHe\ in Gelsil glass,
concerning the experimentally studied temperature range.
In fact, as we will demonstrate in the following,
a large part of the experimental data fits
quite well with the simple 4D XY scaling.
At very low temperatures, we see a crossover
from the 4D XY behavior, which would be primarily
due to the $\delta r$ perturbation.

\begin{figure}[h]
   \centering
\includegraphics[width=0.35 \textwidth]{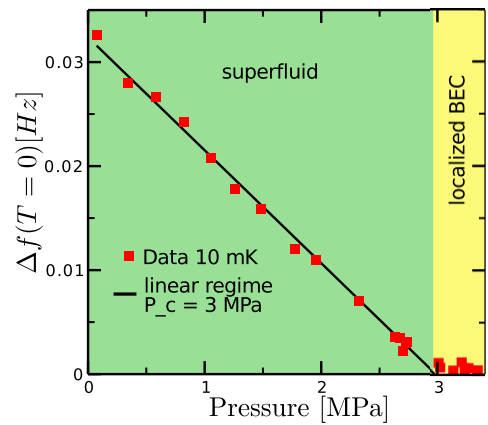}
\caption{
(color online)
Frequency shift in the torsional oscillator experiment,
which is proportional to the superfluid density,
extrapolated to zero temperature and as a function of the pressure.
The experimental data, shown in squares, are taken from Ref ~\cite{shira_prl}.
The solid line is the 4D XY linear scaling~\protect\eqref{eq:zerot}.
The agreement between the data and the scaling is remarkable
except in the close vicinity of the quantum critical point.}
\label{fig:linear_rho}
\end{figure}

\emph{Superfluid density at zero temperature--}
Let us discuss a few consequences of the 4D XY critical
behavior, which we compare with the experiments.
The superfluid density scales with the exponent $\zeta = (d+z-2)\nu$
at zero temperature.
For the 4D XY universality class, $d=3, z=1$ and
the correlation length exponent is $\nu_{4D}=1/2$.
Thus we find
\begin{equation}
\rho_s (p, T=0)\propto (p_c(0)-p)^{\zeta_{4D}}= p_c(0) - p,
\label{eq:zerot}
\end{equation}
where $p_c(0)$ is the critical pressure at zero temperature.
We show the experimental data
for the superfluid density extrapolated to zero temperature
in Fig.~\ref{fig:linear_rho}, and compare with
the prediction~(\ref{eq:zerot}).
The excellent agreement of the data with the linear scaling
strongly supports our proposal based on the 4D XY universality class.
For example, if we had $z=2$ instead of $z=1$,
it would follow $\rho_s \propto (p_c(0)-p)^{3/2}$ which is
clearly inconsistent with the data.

We note that the extrapolation of the linear scaling
to zero temperature gives $p_c(0) \sim 3.0$ MPa
instead of the $3.4$ MPa which is the critical
pressure directly obtained from the experimental data.
We suppose that the non-random 4D XY scaling is valid
for the effective critical pressure
$p^{\rm eff}_c(0) \sim 3.0$ MPa;
this is different from the ``true''
critical pressure $p_c(0)$, which is
likely to be affected by the randomness.

\begin{figure}[h]
   \centering
\includegraphics[width=0.35 \textwidth]{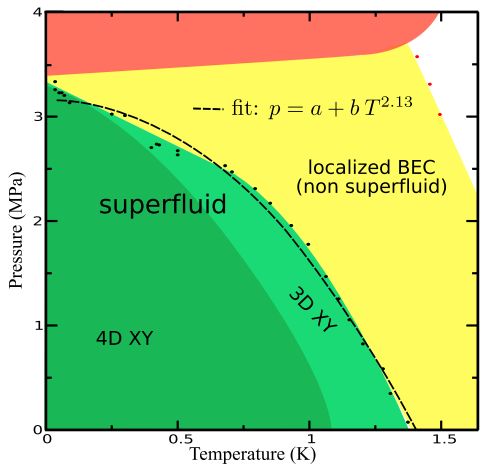}
\caption{ (color online) Phase diagram as obtained in \cite{shira_prl} and the fit to the experimental data.}
\label{pic:fit}
\end{figure}

\emph{Finite temperatures--}
At finite temperature $T$, the length of the system
in imaginary time direction takes the finite value $1/T$
and we impose periodic boundary conditions 
in this direction, see for example \cite{sondhi}.
The finite temperature problem is thus equivalent to the classical
XY model on a \emph{hyperstrip} geometry where the base is an
infinite three dimensional simple cubic lattice and the
\emph{slab width} is given by $1/T$.
The system at finite temperature is thus amenable to a
finite size scaling analysis~\cite{barber}
in the finite imaginary time direction.
This allows us to make several predictions on the
temperature dependence of the system.

For example, the critical pressure $p_c(T)$ at temperature $T$ follows
\begin{equation}
p^{\rm eff}_c(0)-p_c(T) \propto T^{1/\nu_{4D}} = T^2.
\label{eq:boundary}
\end{equation}
In Fig.~\ref{pic:fit} we fitted a power-law to the experimental data
of the phase boundary 
between the superfluid and the non-superfluid (LBEC) phases
presented in \cite{shira_prl}.
We find an exponent of 2.13 which
is in good agreement with equation \eqref{eq:boundary},
while $p^{\rm eff}_c(0) \sim 3.2$ MPa is slightly
larger than that obtained in Fig.~\ref{fig:linear_rho}.

\emph{Discussion--}
We have demonstrated that the experimental data
on \FourHe\ in Gelsil glass agrees with 4D XY scaling
rather well.
This corresponds to
the theory~\eqref{eq:Lag} with a small perturbation $\delta r$,
displaying clean 4D XY quantum critical behavior
sufficiently far away from the quantum
critical point. 
However, close to the quantum critical point,
the effect of $\delta r$ is enhanced and
crossover to a different universality class should occur.
In fact, by inspection of Fig.~\ref{pic:fit}, the phase boundary
at very low temperatures
starts to deviate in a pronounced way from the
4D XY scaling~\eqref{eq:boundary}.
We also find that, in Fig.~\ref{fig:linear_rho},
the superfluid density at zero temperature
also deviates from the 4D XY scaling~\eqref{eq:zerot},
close to the critical pressure $p_c(0)$.
These effects of the disorder are the subject of future investigations. 

We also note that, close to the quantum critical point,
we find that the extraction of physical quantities
(such as the superfluid density and the critical temperature)
from the raw frequency data also becomes rather subtle.
A more precise analysis would require a more sophisticated
analysis of the raw data based on a better theoretical understanding.
In any case, we believe that the analyses presented in this Letter
demonstrate the basic validity of the  proposed picture.

We thank Igor Herbut for useful discussions.
K.~S. is supported by 
Grant-in-Aid for Scientific Research on Priority Area
"Physics of new quantum phases in superclean materials"
(Grant No. 17071010), and T.~E. was supported by
a Government Scholarship, both from MEXT of Japan.

\end{document}